\documentclass[aps,prb,reprint,superscriptaddress]{revtex4-2}

\usepackage{color}
\usepackage[dvipdfmx]{graphicx}

\usepackage{amsmath}
\usepackage{graphicx}
\usepackage{dcolumn}
\usepackage{bm}
\usepackage{txfonts}

\begin{document}

\title{Interband spectroscopy of Landau levels and 
magnetoexcitons in bulk black phosphorus}

\author{H.~Okamura}
\altaffiliation[Electronic address: ]{ho@tokushima-u.ac.jp}
\affiliation{Department of Applied Chemistry, 
Tokushima University, Tokushima 770-8506, Japan}

\author{S. Iguchi}
\author{T. Sasaki}
\affiliation{Institute for Materials Research, 
Tohoku University, Sendai 980-8577, Japan}

\author{Y. Ikemoto}
\author{T. Moriwaki}
\affiliation{Japan Synchrotron Radiation Research 
Institute, Sayo 679-5198, Japan}

\author{Y. Akahama}
\affiliation{Graduate School of Materials Science, 
University of Hyogo, Hyogo 678-1297, Japan}
\date{\today}

\begin{abstract}
Low-dimensional systems based on black phosphorus 
(BP) have recently attracted much interest.  
Since the crystal structure of BP consists of 2D 
layers with a strong in-plane anisotropy, its 
electronic properties at a high magnetic field 
($B$) are quite interesting.   
Here, we report a magneto-optical study of bulk BP 
at high $B$ to 12~T perpendicular to the 2D layers.  
In the obtained optical conductivity spectra, 
periodic peaks are clearly observed corresponding 
to Landau levels of up to $n=6$ quantum number.  
They exhibit almost linear shifts with $B$, and 
from their analysis an electron-hole reduced mass 
of 0.13~$m_0$ is obtained, where $m_0$ is the 
electron mass.  
Many of the peaks appear in pairs, which is 
interpreted in terms of Zeeman splitting and 
an electron-hole combined $g$-factor of 6.3 
is obtained.     
In addition, a magnetoexciton peak is observed to 
shift quadratically with $B$, and the exciton binding 
energy at $B=0$ is estimated to be 9.7~meV.  
\end{abstract}

\maketitle
\section{Introduction}
Black phosphorus (BP) is one of the allotropes of 
phosphorus, which has a layered crystal structure 
with orthorhombic symmetry at ambient condition 
\cite{structure,footnote}.   
As illustrated in Fig.~1(a), the structure consists 
of layers of P atoms formed by covalent bonding.  
The P atoms form a puckered, or the so-called 
``arm chair", pattern along the $x$ axis and a 
zigzag pattern along the $y$ axis.  The layers are 
stacked along the $z$ axis by van der Waals bonding 
with a spacing of 5.3~\AA.  
Although BP was first reported in 1914 \cite{bridgman}, 
it was only after the successful synthesis of large 
single crystals in the 1980s \cite{shirotani,endo} 
that its physical properties were fully investigated 
\cite{akahama-transport,narita,akahama-pressure1,
SC,morita,akahama-pressure2,akahama-optical}.   
At ambient condition, BP is a narrow-gap semiconductor 
with a band gap of 0.33~eV.  According to band 
calculations \cite{morita-1,morita-2,newbandcalc} the 
minimum band gap is a direct gap located at the $Z$ 
point of the Brillouin zone.  
An optical transition across the band gap is dipole 
allowed for polarization along the $x$ axis but forbidden 
for that along the $y$ axis \cite{morita-2,newbandcalc}, 
resulting in large in-plane optical anisotropies as 
confirmed by experiment \cite{anisotropy}.  
Recently, a successful realization of phosphorene 
\cite{phosphorene1,phosphorene2}, namely an 
atomically thin layer of exfoliated BP, has led 
to a tremendous amount of research on BP-based 
low-dimensional systems \cite{phosphorene3}.  
In particular, the ability to tune the band gap 
\cite{newbandcalc,gap-control-2,gap-control-3} 
with the layer number makes phosphorene quite 
attractive for practical device applications 
\cite{phosphorene3}.  
Another remarkable feature of BP is a rich 
variety of phenomena induced by applying 
a high magnetic field 
\cite{BP-QHE,Zhou2015-1,Zhou2015-2,Peeters,
Low,Pereira,Zhou2017,Zhou2023,pump-probe}, 
a high pressure 
\cite{akahama-pressure1,SC,perucchi1,layer-pressure,goodenough}, 
and both of them \cite{akiba,xiang,mito}.  
The present work focuses on the electronic properties 
of bulk BP under high magnetic field ($B$) probed by 
interband optical spectroscopy.

The electrons at high $B$ undergo a cyclotron motion 
and their energy spectrum is quantized into discrete 
Landau levels (LLs).   In a 2D electronic system 
this Landau quantization leads to the quantum 
Hall effect in transport properties \cite{QHE}, 
which was initially studied in semiconductor 
heterostructures and later in other 2D systems such 
as graphene \cite{graphene-QHE-1,graphene-QHE-2} 
and phosphorene \cite{BP-QHE}.  
The quantization may also be directly observed 
on the energy axis by a magneto-optical study, 
either by an intraband spectroscopy such as 
cyclotron resonance \cite{CR} or by an interband 
spectroscopy such as optical absorption/reflection 
and photoluminescence \cite{heiman}.  
The properties of LLs and cyclotron motion in 
the highly anisotropic 2D layers of BP are quite 
interesting, and have motivated a large number of 
theoretical studies \cite{Zhou2015-1,Zhou2015-2,
Peeters,Low,Pereira,Zhou2017,Zhou2023}.  
For example, unconventional $B$ dependences of 
LL energies and optical selection rules have been 
predicted for few-layer and thin-film BPs 
\cite{Zhou2015-2,Peeters}.    
Rather surprisingly, however, there have been much 
fewer experimental reports on the magneto-optical 
properties of BP \cite{narita,pump-probe} compared 
with the theoretical ones.   
For example, cyclotron resonance studies of bulk BP 
\cite{narita} were made soon after large 
single crystals became available.   
More recently, an ultrafast laser spectroscopy of 
bulk BP was made to study effects of magnetic fields 
on the anisotropic electron dynamics \cite{pump-probe}.  
Nevertheless, magneto-optical properties of BP 
remain largely unexplored experimentally, and 
providing more magneto-optical data on BP should 
be essential for further development of BP research.

In this work, we have used an interband 
magnetoreflection technique to probe 
the intriguing electronic properties of bulk 
BP at high $B$ perpendicular to the 2D layers.  
In this technique, the optical reflectivity spectrum 
[$R(\omega)$] of a sample measured at high $B$ 
exhibits periodic peaks corresponding to 
interband transitions between LLs in valence and 
conduction bands.  
Alternatively, $R(\omega)$ at a fixed $\omega$ 
may be measured as a function of $B$.  
This technique has been applied to study LLs in 
various narrow-gap semiconductors and semimetals 
including Bi \cite{Bi-1,Bi-2} and graphite 
\cite{graphite1,graphite2}.   
Similarly, interband absorption has also been 
measured to probe, for example, LLs and Dirac 
states in graphene \cite{graphene-LL}.  
An advantage of these interband techniques is that 
the LLs can be observed even in the absence of 
excess carriers.  
%
We have clearly observed periodic peaks due to 
LL formation in $R(\omega)$ of bulk BP at $B \geq 5$~T, 
which shift almost linearly with $B$.   
The peak shifts seem to show conventional $B$ 
dependence in spite of the large in-plane anisotropy.  
From the data, the electron-hole reduced mass and 
effective $g$-factor are obtained.    
In addition, we have clearly identified a magnetoexciton 
peak and the exciton binding energy is obtained.    
\begin{figure}
\begin{center}
\includegraphics[width=0.45\textwidth]{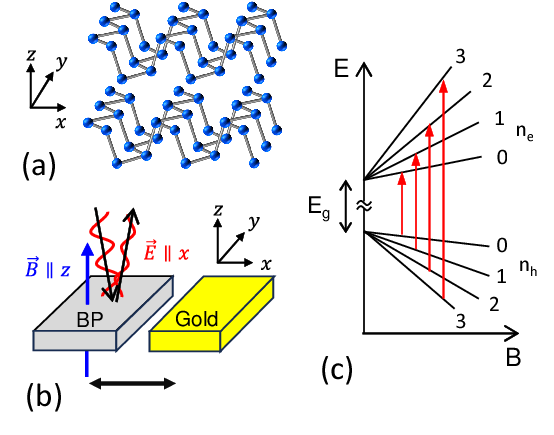}
\end{center}
\caption{
(a) Crystal structure of BP at ambient condition 
\cite{structure,footnote}.   The $x$ axis is along 
the so-called ``arm chair" direction.   
(b) The configuration of sample and gold 
reference in the magnetoreflectivity study.   
External magnetic field ($\vec{B}$) is applied 
along the $z$ axis and the incident light is 
polarized ($\vec{E}$) along the $x$ axis.  
(c) A schematic diagram for LL formation as 
discussed in the text.  $E$ represents the 
electron energy, and $E_g$ the band gap.  
(For simplicity, Zeeman splitting is omitted 
here.)  
The red arrows indicate interband optical 
transitions between LLs with the same quantum 
number, $n_e=n_h$.
}
\end{figure}

\section{Experimental Methods}
The samples of BP used in this work 
were undoped single crystals \cite{carrier} 
with a thickness of about 100~$\mu$m grown 
by a high-pressure synthesis as described 
elsewhere \cite{endo}.  
A freshly cleaved sample surface containing 
the $xy$ plane was used.     
$R(\omega)$ spectra of the sample at high $B$ 
were measured at the magneto-optics 
endstation \cite{MO} of the infrared beamline 
BL43IR at SPring-8 \cite{bl43ir} using 
synchrotron radiation as a bright infrared 
source \cite{IRSR-review}.   Magnetic fields 
up to 12~T were generated by a superconducting 
magnet equipped with an infrared microscope.  
The diameter of the focused infrared beam 
on the sample was smaller than 50 $\mu$m, and 
a flat portion of the sample was chosen for 
the measurement.
A BP sample and a gold reference mirror were 
mounted on the cold finger of a liquid He 
continuous-flow cryostat with a KBr optical 
window, which was inserted into the magnet bore.   
At each $B$, a measured reflection spectrum of 
the sample was divided by that of the gold to 
obtain the reflectivity, $R(\omega)$.   The 
spectra were recorded with an FTIR spectrometer 
and a HgCdTe detector. 
As illustrated in Fig.~1(b), the measurements 
were made under a Faraday geometry, where 
$\vec{B}$ was applied perpendicular to the sample 
surface ($xy$ plane) and $R(\omega)$ was measured 
under a near-normal incidence.   
The incident light was polarized along the $x$ axis, 
which leads to dipole-allowed direct transitions 
across the band gap.

\section{Results and Discussions}
Figure~2 indicates the optical conductivity 
[$\sigma(\omega)$] of bulk BP measured 
at 12~K and at different $B$ to 12~T 
over a spectral range in which the band gap 
is located.
\begin{figure}
\begin{center}
\includegraphics[width=0.46\textwidth]{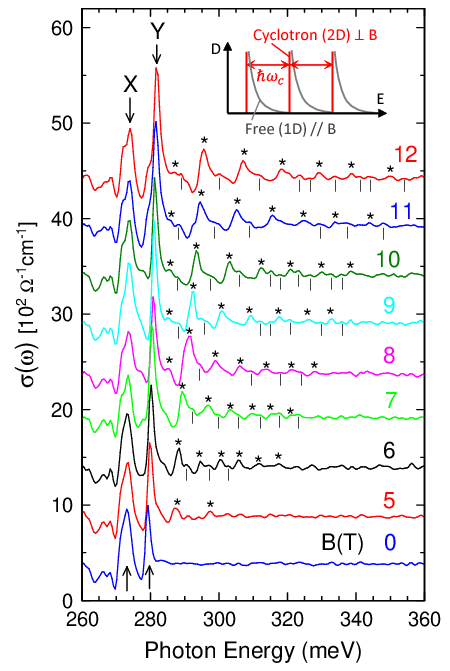}
\end{center}
\caption{
Optical conductivity [$\sigma(\omega)$] spectra of 
bulk BP measured at 12~K under high magnetic 
field ($B$) exhibiting periodic peaks due to LL 
formation.   
Each spectrum is vertically offset 
by 500~$\Omega^{-1}$cm$^{-1}$ for clarity.  
The asterisks and vertical bars indicate main peaks 
and subpeaks, respectively, discussed in the text.  
The vertical arrows indicate 
an exciton peak (X) and a bound state peak (Y). 
The inset illustrates the electron density of 
states ($D$) versus energy ($E$) at high $B$. 
$\hbar\omega_c$ denotes the cyclotron energy.  
}
\end{figure}
They have been obtained from the measured $R(\omega)$ 
spectra, which are indicated in Supplemental 
Material \cite{SI}, 
using the Kramers-Kronig analysis \cite{mybook,footnote2}.  
It is seen that the spectral shapes of $\sigma(\omega)$ 
are very similar to those of the corresponding $R(\omega)$. 
At $B=0$, the main spectral features in 
$\sigma(\omega)$ are the two peaks indicated 
by vertical arrows, which are located at 
273.0 and 279.3~meV and labeled as X and Y, 
respectively.  
Later, it will be shown that the 
X peak corresponds to the exciton ground state 
and the Y peak to a bound electron or hole state.  
Except for these two peaks, the 
spectra are almost featureless at $B=0$.  
At $B \geq$ 5~T, however, multiple peaks emerge 
as marked with the asterisks and vertical bars in Fig.~2. 
Note that the peaks with asterisks, which we 
hereafter refer to as the ``main peaks", appear 
periodic in energy and their energy interval 
increases with $B$.   On the other hand, the 
peaks with vertical bars, which we refer to as 
the ``subpeaks", are weaker and less clearly 
observed than the main peaks.     These periodic 
peaks are apparently associated with interband 
optical transitions between LLs formed in 
valence and conduction bands, as shown by 
Fig.~1(c) \cite{heiman}.   
The optical transitions in Fig.~1(c) are 
indicated with the conventional selection rule 
(between LLs with the same quantum number), 
although some unconventional selection rules 
have been predicted for a few-layer BP 
\cite{Zhou2015-2} as discussed later.  
Note that bulk BP is not a completely 2D system 
since the interlayer electronic coupling is fairly 
strong \cite{morita-1,morita-2,newbandcalc}.  
Therefore, the energy dispersion along the $z$ 
axis is continuous even when that along the $xy$ 
plane is quantized by a magnetic field. 
Despite this, the measured $\sigma(\omega)$ still 
exhibits a LL quantization because, for the 1D 
motion of electrons along $z$, 
the density of states ($D$) decreases with 
energy ($E$) as $D \propto 1/\sqrt{E}$ 
as shown by the inset of Fig.~2.

To further analyze the data, the overall peak 
energies of $\sigma(\omega)$ in Fig.~2 are 
plotted versus $B$ in Fig.~3(a), where the main 
peaks and subpeaks are indicated by filled 
and empty circles, respectively.  
%
\begin{figure}
\begin{center}
\includegraphics[width=0.45\textwidth]{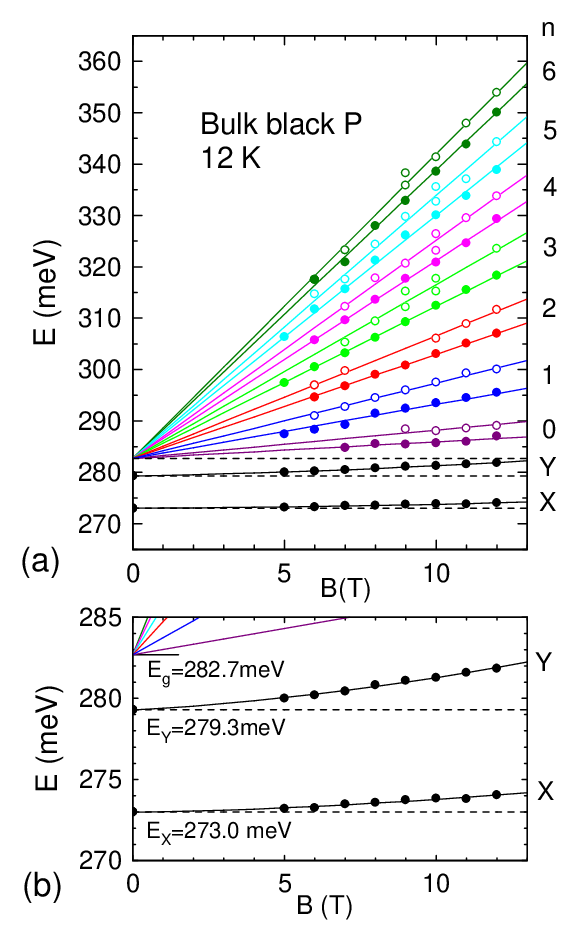}
\end{center}
\caption{(a) The energies ($E$) of main peaks 
(colored filled circles) and subpeaks (empty 
circles), which are marked by asterisks and 
vertical bars in Fig.~2, respectively, and 
X and Y peaks (black filled circles) 
plotted versus $B$.  
The colored lines indicate results of linear 
fitting to the peaks with $n=0$$-$6 and the 
black curves those of quadratic fitting to 
X and Y peaks.   The horizontal broken lines 
are guide to the eye.  
(b) The $B$ dependences of X and Y peaks 
displayed on an expanded scale.  The band gap 
($E_g$) estimated from the $n=0$$-$6 peaks 
and the observed peak energies of X and Y 
at $B=0$ ($E_\text{X}$ and $E_\text{Y}$) 
are also indicated.}
\end{figure}
This so-called ``fan plot" clearly shows the 
characteristic variation of transition energy 
with $B$.   
Pairs of main peak and subpeak, marked by 
$n=0$ to 6, are observed to shift almost linearly 
with $B$.  In addition, the interval between the 
main peak and subpeak is also increasing with $B$.  
These peaks exhibit much larger shifts with $B$ 
than X and Y peaks, which suggests different 
characteristics between the two groups of peaks.   
The former group of peaks should arise from free 
electron-hole ($e-h$) pair creations and their linear $B$ 
dependence is due to the increase of cyclotron 
energy.  
In a simple free-electron model, the LL energy spectra 
of $e$ and $h$ are expressed as 
$\epsilon_{n,i}=(n_i+1/2)\hbar\omega_{c,i}$, 
where $n_i=0, 1, 2, \cdots$ is the quantum number, 
$\omega_{c,i}$ is the cyclotron energy, and 
$i=e$ or $h$ \cite{heiman,miura}.   
As already mentioned, interband transitions are 
allowed only between LLs with $n_e=n_h$, so the 
interband transition energy is expressed as 
%
\begin{equation}
E_n = E_g + \left(n + \frac{1}{2} \right) \hbar \omega_c^\ast
 =E_g + \left(n + \frac{1}{2} \right) 
\left( \frac{\hbar e}{m_r^\ast} \right) B.
\end{equation}
%
Here, $E_g$ is the band gap, $n=0,1,2,\cdots$, 
$\omega_c^\ast = \omega_{c,e} + \omega_{c,h}$ is 
the combined cyclotron frequency, 
and $1/m_r^\ast = 1/m_e^\ast + 1/m_h^\ast$ where 
$m_r^\ast$ is the reduced effective mass, and 
$m_e^\ast$ ($m_h^\ast$) is the electron (hole) 
effective mass.   
Namely, the transition energy is expected to 
shift linearly with $B$ and its slope should be 
given by $\left( n+ \frac{1}{2} \right) \hbar e/m_r^\ast$.  
Accordingly, a linear fitting of the form 
$E=E_g + S \cdot B$ has been performed on 
the data in Fig.~3(a), where the slope $S$ 
is a fitting parameter and $E_g$ (the vertical 
intercept of the graph) is kept the same for 
all the peaks.   
It has been found that the best overall fitting 
is obtained when $E_g$ is in the range $282.7 \pm 0.5$~meV.  
The result of fitting with $E_g=282.7$~meV is 
indicated by the colored straight lines in 
Fig.~3(a).
Note that the band gap of bulk BP at 12~K obtained 
here, $E_g=282.7$~meV, is much smaller than that 
at room temperature (330~meV) since $E_g$ of 
bulk BP is significantly reduced by cooling 
\cite{gap-Tdep}.

The slopes obtained by the fitting in Fig.~3 
are summarized in Table~I and plotted versus 
$n$ in Fig.~4.  
%
\begin{table}[t]
\caption{
Slopes ($S_n$) in units of meV/T obtained 
from the linear fitting to the peaks with 
$n=0$$-$6 and $E_g=282.7$~meV in Fig.~3(a).  
$S_n$(main) and $S_n$(sub) indicate the slopes 
of the main peak and subpeak, which are marked 
by the asterisks and vertical bars in Fig.~2, 
respectively. 
$S_n$(mid) and $\Delta S_n$ indicate the middle 
value of $S_n$(main) and $S_n$(sub) and the 
difference between them, respectively.  
The variation in $S_n$ when $E_g$ is varied by 
$\pm$~0.5~meV is about $\pm$~0.5~meV for the main 
peaks for all $n$.}
\begin{ruledtabular}
\begin{tabular}{cccccccc}
$n$ & 0 & 1 & 2 & 3 & 4 & 5 & 6 \\
\colrule
$S_n$(main) & 0.322 & 1.05 & 2.03 & 2.96 & 3.85 & 4.73 & 5.61 \\
$S_n$(sub) & 0.55 & 1.47 & 2.39 & 3.38 & 4.25 & 5.12 & 5.93 \\
$S_n$(mid) & 0.436 & 1.26 & 2.21 & 3.17 & 4.05 & 4.93 & 5.77 \\
$\Delta S_n$  & 0.23 & 0.42 & 0.36 & 0.42 & 0.39 & 0.39 & 0.32 \\
\end{tabular}
\end{ruledtabular}
\end{table}
%
\begin{figure}
\begin{center}
\includegraphics[width=0.42\textwidth]{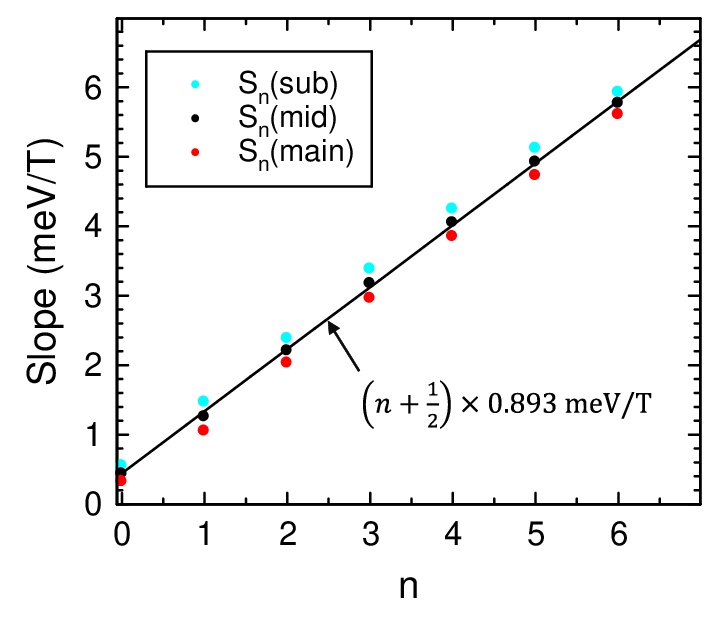}
\end{center}
\caption{
Slopes of the main peak [$S_n$(main), red dots], 
sub-peak [$S_n$(sub), blue dots] and their middle 
value [$S_n$(mid), black dots] from Table~I are 
plotted versus $n$.  
The solid line shows a linear fit to $S_n$(mid). 
}
\end{figure}
Since many of the peaks appear in pairs, we assume 
that each pair of adjacent main peak and subpeak 
results from the Zeeman splitting of a LL.  
Theoretically, more complicated Zeeman splitting may be 
expected since the wavefunction around $Z$ point 
should contain a strong $p$ component 
\cite{morita-1,morita-2}.  Namely, a spin-orbit coupling 
may split the $p$ states into $j=3/2$ and 1/2, which may 
further split into $m_j=\pm3/2$ and $\pm1/2$ states 
at high $B$, where $j$ and $m_j$ are total angular momentum 
and magnetic quantum numbers, respectively \cite{heiman,miura}.  
In our data, however, such complicated splitting 
is not observed \cite{splitting}.  
We therefore assume phenomenologically a Zeeman 
splitting of the form 
\begin{equation}
E_n^\pm = E_n \pm \frac{1}{2} \Delta E
\end{equation}
where $\Delta E= g^\ast \mu_\text{B} B$ and 
$g^\ast$ is the $e$-$h$ combined effective $g$-factor, 
$\mu_\text{B}$ is the Bohr magneton.  We assume that 
$E_n^-$ and $E_n^+$ correspond to 
the main peak and subpeak, respectively.  
Then, the middle value of $S_n$(main) and $S_n$(sub) 
should correspond to the slope of the $n$th LL.  
In Table~I and Fig.~4, the middle values are 
indicated and denoted as $S_n$(mid).   
As indicated by the straight line in Fig.~4, 
$S_n$(mid) can be fitted well as 
$S_n$(mid)$=\left(n+\frac{1}{2} \right) 
\times 0.893$~meV/T.   Namely, the $B$ dependence 
of $S_n$(mid) is consistent with Eq.~(1).  
It is unclear why the subpeak is much weaker 
than the main peak.  The intensity of a peak in 
$\sigma(\omega)$ should be proportional to the joint 
density of states (JDOS) and the transition probability ($P$).  
In our simplified model, the main peak and subpeak 
correspond to interband transitions 
$(\epsilon^+_{n,h} \rightarrow \epsilon^-_{n,e})$ 
and 
$(\epsilon^-_{n,h} \rightarrow \epsilon^+_{n,e})$, 
respectively, where $-$ and $+$ symbols express 
the lower- and higher-energy Zeeman split levels, 
respectively \cite{footnote4}.  
Both JDOS and $P$ should be the same for the 
two transitions.   
Therefore, other factors not considered here, 
such as a mixing among spin-orbit split states 
at high $B$, may have resulted in their different 
intensities.  

The LL peak energies in Fig.~3(a) seem to follow 
a conventional $B$ dependence of Eq.~(1), despite 
the large in-plane anisotropy in BP.   
In the case of thin-film BPs, in contrast, unconventional 
selection rules such as $n_e = n_h \pm 2$ and nonlinear 
shifts of LL peaks with $B$ have been predicted 
as already mentioned \cite{Zhou2015-2}.   
Theoretically, similar results may be expected for 
a bulk BP at certain values of $k_z$, but again, 
it may not be observed in actual experiments 
\cite{Zhou2024}.  
Comparing $S_n$(mid)$=\left(n +\frac{1}{2} \right) 
\times 0.893$~meV/T 
with Eq.~(1), 0.893~meV/T should be equal to 
$(\hbar e/m_r^\ast)$, which yields 
$m_r^\ast=0.129 m_0$. 
Previous cyclotron resonance study of $n$-type 
and $p$-type BP reported in-plane effective 
masses of $m_e^\ast=0.291 m_0$ 
and $m_h^\ast=0.222 m_0$ \cite{narita}, 
which give a reduced effective mass of 
$m_r^\ast=0.126 m_0$.  
Therefore, the present result of 
$m_r^\ast=0.129 m_0$ is very close to the 
cyclotron resonance result.  
%
In our model with Eq.~(2), 
$\Delta S_n = S_n$(sub)$-S_n$(main) should 
be equal to $\Delta E / B$, and hence equal to 
$\left| g^\ast \right| \mu_\text{B}$.  
(Note that the sign of $g^\ast$ cannot be 
determined in our study.)  
This relation and the weighted average of $\Delta S_n$ 
from Table~I, 0.37~meV/T, yields 
$\left| g^\ast \right|= 6.3$.
A theoretical study has estimated the electron 
effective $g$-factor in bulk BP to be 3 \cite{Zhou2017}, 
which is about one-half the present result.  
Since $\left| g^\ast \right|$ obtained here contains both 
electron and hole contributions, our result of 
$\left| g^\ast \right|=6.3$ seems reasonable in 
comparison with the theoretical estimate.    
Note that the $n=0$ LL peaks in $\sigma(\omega)$ are 
observed much less clearly than the $n=1$$-$3 peaks.  
The reason for this result is unclear, but note also 
that the $n=0$ LL peak is not always stronger than 
$n \geq 1$ peaks in reported results on other materials, 
and $n=0$ peak is sometimes not observed even when 
$n \geq 1$ peaks are clearly observed \cite{perovskite}.  
The weaker $n=0$ peak in the present study may have resulted 
from its proximity to the band edge, but its microscopic 
origin is unclear at present.

Below we look more closely into the $B$ dependences 
of X and Y peak energies, which are displayed 
in an expanded scale in Fig.~3(b).   
The solid curves in Fig.~3(b) indicate results 
of quadratic fitting of the form 
$E(B)=aB^2 + bB + E_\text{X,Y}$ to the 
data, where $a$ and $b$ are fitting parameters.  
The $E(B)$ obtained with the fitting \cite{footnote3} 
has a strong $B^2$ term comparable with the $B$ term.  
Such a quadratic shift with $B$ is a well-known 
property of 
bound electron and 
exciton states \cite{heiman,miura,diamagnetic} and 
is often referred to as a diamagnetic shift.  
X peak was clearly observed at elevated temperatures 
to room temperature, and was observed even for samples 
not freshly cleaved (not shown).  From these results, 
we attribute X peak to absorption by intrinsic (free) 
exciton ground (1$s$) state. 
As discussed above, the results on LL peaks have 
yielded $E_g = 282.7 \pm 0.5$~meV.  
The exciton binding energy $E_b$ should be given 
by the difference between $E_g$ and $E_\text{X}$ 
at $B=0$, so we obtain $E_b = 9.7 \pm 0.5$~meV.  
This is close to a theoretical estimate of 9.1~meV 
\cite{BP-exciton} obtained by a fully anisotropic 
exciton model. 
From the measured $E_b=9.7$~meV, the exciton Bohr 
radius can be roughly estimated to be $a^\ast = 55~\AA$ 
using the isotropic hydrogenic model and 
$m_r^\ast=0.129 m_0$ obtained above \cite{radius}.  
For comparison, the cyclotron radius is 
$r_c=(\hbar/eB)^{1/2}=81~\AA$ at $B$=10~T \cite{miura}.  
%
As for Y peak, the binding energy relative to 
$E_g$ is 3.4~meV from Fig.~3(b), and it exhibits 
much larger shifts with $B$ than the X peak.  
Such results are often observed for excited (2$s$) 
state of excitons \cite{miura,diamagnetic}.  
However, it is also well known that a 2$s$ exciton 
absorption is usually much weaker than the 
corresponding 1$s$ exciton one \cite{exciton}.  
This is in contrast to the present data in Fig.~2, 
where at $B=0$ the intensity of Y peak is not much 
lower than that of X peak.  Therefore, Y peak 
cannot be attributed to 2$s$ exciton peak here.   
An earlier study on the reflectivity of bulk BP at 
zero field \cite{morita}, which was also performed 
on a high-pressure synthesized single crystal, reported 
a pair of peaks near $E_g$ which appear very analogous 
to X and Y peaks in our data.  
They interpreted the higher-energy peak (Y peak 
in our case) to be of extrinsic origin since it 
strongly depended on experimental conditions such 
as the time elapsed after cleavage and the degree 
of vacuum during measurements.  
In our study, Y peak was observed for freshly 
cleaved samples only.  From these results, we 
interpret Y peak as arising from some trapped 
electron or hole state likely caused by cleavage.  
Then, the strong diamagnetic shifts of Y peak is 
reasonable since it is well known that a trapped 
electron state also exhibits quadratic shifts 
with $B$ similarly to excitons \cite{miura}.  
The exact microscopic nature of the trapped state 
givng rise to Y peak is, however, unclear 
at present and should be addressed further 
in future studies.

Note that our clear observation of 
magnetoexcitons in a bulk BP makes similar studies 
on few-layer and 
thin-film BPs quite appealing.  Since the exciton 
binding energy should be much more enhanced than in 
the bulk \cite{gap-control-2,gap-control-3}, 
magnetoexciton properties in a few-layer BP are 
quite intriguing.   
In particular, even more complex exciton states such 
as trions ({\it e.g.}, a charged exciton consisting 
of two electrons and a hole \cite{okamura2}) may 
be explored in few-layer BPs.

\section{Summary}
Infrared magnetoreflection study has been made on 
bulk BP single crystals at high $B$ perpendicular 
to the 2D layers.   
At $B \geq$ 5~T, the obtained $\sigma(\omega)$ 
spectra exhibit periodic peaks due to interband 
transitions between LLs in conduction and 
valence bands. 
The peaks show linear shifts with $B$, which agree 
well with the standard form of 
$E_n = E_g + \left( n+\frac{1}{2} \right) \hbar \omega_c^\ast$. 
From the data, an electron-hole reduced mass of 
$m_r^\ast = 0.129 m_0$ is obtained, which agrees 
well with that reported previously by a 
cyclotron resonance study.  
In addition, from the observed interval between 
the main peak and subpeak, an electron-hole combined 
effective $g$-factor is obtained as 
$\left| g^\ast \right|=6.3$. 
Furthermore, a magnetoexciton peak with characteristic 
quadratic-in-$B$ shifts has also been observed, 
and the exciton binding energy at $B=0$ is 
estimated to be 9.7~$\pm$ 0.5~meV.

\begin{acknowledgments}
The authors would like to thank Dr. X. Zhou for 
useful communications.  They also thank K. Nishino, 
T. Goto and A. Tsubouchi for technical assistance.  
The experiments at SPring-8 were performed under 
the approval by JASRI 
(2016A0073, 2017A1164, 2017B1389, 2018A1130). 
Financial support from Japan Society for the Promotion 
of Science (KAKENHI Grant No. 26400358) is acknowledged.  
\end{acknowledgments}



\begin{thebibliography}{99}

\bibitem{structure}
A. Brown and S. Rundqvist, 
Refinement of the crystal structure of black phosphorus, 
Acta Crystallogr. {\bf 19}, 684 (1965).  

\bibitem{footnote}
The $x$, $y$, and $z$ directions here correspond 
to the crystallographic $c$, $a$, and $b$ axes, 
respectively \cite{structure}.  

\bibitem{bridgman} 
P. W. Bridgman,
Two new modificatons of phosphorus, 
J. Am. Chem. Soc. {\bf 36}, 1344 (1914). 

\bibitem{shirotani} 
I. Shirotani, 
Growth of Large Single Crystals of Black Phosphorus, 
Mol. Cryst. Liq. Cryst. {\bf 86} (1982) 203.

\bibitem{endo}
S. Endo, Y. Akahama, S. Terada, and S. Narita, 
Growth of Large Single Crystals of 
Black Phosphorus under High Pressure, 
Jpn. J. Appl. Phys. {\bf 21}, L482 (1982). 

\bibitem{akahama-transport} 
Y. Akahama, S. Endo, and S. Narita, 
Electrical Properties of Black Phosphorus Single Crystals, 
J. Phys. Soc. Jpn. {\bf 52}, 2148 (1983).  

\bibitem{narita}
S. Narita, S. Terada, S. Mori, K. Muro, 
Y. Akahama, and S. Endo, 
Far-Infrared Cyclotron Resonance Absorptions 
in Black Phosphorus Single Crystals, 
J. Phys. Soc. Jpn. {\bf 52}, 3544 (1983).

\bibitem{akahama-pressure1}
M. Okajima, S. Endo, Y. Akahama, and S. Narita, 
Electrical Investigation of Phase Transition in 
Black Phosphorus under High Pressure, 
J. Jpn. Appl. Phys. {\bf 23} 15 (1984).  

\bibitem{SC}
H. Kawamura, I. Shirotani, and K. Tachikawa, 
Anomalous superconductivity in black phosphorus 
under high pressures, 
H. Kawamura, I. Shirotani, and K. Tachikawa, 
Solid State Commun. {\bf 49}, 879 (1984).

\bibitem{morita}
Early experiments performed on large single crystals 
of BP in 1980s have been summarized in A. Morita, 
Semiconducting black phosphorus, 
Appl. Phys. A {\bf 39}, 227 (1986).  


\bibitem{akahama-pressure2} 
Y. Akahama and S. Endo, 
Transport study on pressure-induced band overlapped 
metallization of layered semiconductor black phosphorus, 
Solid State Commun. {\bf 104}, 307 (1997). 

\bibitem{akahama-optical}
Y. Akahama and H. Kawamura, 
Optical and Electrical Studies on Band-Overlapped 
Metallization of the Narrow-Gap Semiconductor Black 
Phosphorus with Layered Structure, 
physica status solidi (b) {\bf 223}, 349 (2001).
\bibitem{morita-1}
Y. Takao, H. Asahina, and A. Morita, 
Electronic Structure of Black Phosphorus in 
Tight Binding Approach, 
J. Phys. Soc. Jpn. {\bf 50}, 3362 (1981).   

\bibitem{morita-2}
H Asahina, K. Shindo, and A. Morita, 
Electronic Structure of Black Phosphorus in 
Self-Consistent Pseudopotential Approach, 
J. Phys. Soc. Jpn. {\bf 51}, 1193 (1982). 

\bibitem{newbandcalc}
J. Qiao, X. Kong, Z.-X. Hu, F. Yang, and W. Ji, 
High-mobility transport anisotropy and linear 
dichroism in few-layer black phosphorus, 
Nature Commun. {\bf 5}, 4475 (2014).  


\bibitem{anisotropy}
F. Xia, H. Wang, and Y. Jia, 
Rediscovering black phosphorus as an anisotropic 
layered material for optoelectronics and electronics, 
Nature Commun. {\bf 5}, 4458 (2014).  

\bibitem{phosphorene1}
L. Li, Y. Yu, G. J. Ye, Q. Ge, X. Ou, H. Wu, 
D. Feng, X. H. Chen, and Y. Zhang, 
Black phosphorus field-effect transistors, 
Nature Nanotech. {\bf 9}, 372 (2014).  


\bibitem{phosphorene2}
H. Liu, A. T. Neal, Z. Zhu, Z. Luo, X. Xu, 
D. Tomanek, and P. D. Ye, 
Phosphorene: An Unexplored 2D Semiconductor 
with a High Hole Mobility, 
ACS Nano {\bf 8}, 4033 (2014). 

\bibitem{phosphorene3} 
For a review on the phosphorene research, 
see, for example, 
J. Miao, L. Zhang, and C. Wang, 
Black phosphorus electronic and optoelectronic 
devices, 2D Mater. {\bf 6}, 032003 (2019).


\bibitem{gap-control-2}
L. Li, J. Kim, C. Jin, G. J. Ye, D. Y. Qiu, 
F. H. da Jornada, Z. Shi, L. Chen, Z. Zhang, 
F. Yang, K. Watanabe, T. Taniguchi, W. Ren, 
S. G. Louie, X. H. Chen, Y. Zhang, and F. Wang, 
Direct observation of the layer-dependent 
electronic structure in phosphorene, 
Nature Nanotech., {\bf 12}, 21 (2017). 

\bibitem{gap-control-3}
G. Zhang, S. Huang, A. Chaves, C. Song, V. O. Ozcelik, 
T. Low, and H. Yan, 
Infrared fingerprints of few-layer black phosphorus, 
Nature Commun. {\bf 8}, 14071 (2017). 



\bibitem{BP-QHE}
L. Li, F. Yang, G.J. Ye, Z. Zhang, Z. Zhu, W. Lou, 
X.-Y. Zhou, Liang Li, K. Watanabe, T. Taniguchi, 
K. Chang, Y. Wang, X. Chen, and Y. Zhang, 
Quantum Hall effect in black phosphorus two-dimensional 
electron system, 
Nature Nanotech. {\bf 11}, 593 (2016). 

\bibitem{Zhou2015-1}
X. Zhou, R. Zhang, J. P. Sun, Y. L. Zou, D. Zhang, 
W. K. Lou, F. Cheng, G. H. Zhou, F. Zhai, and Kai Chang, 
Landau levels and magneto-transport property of 
monolayer phosphorene, 
Sci. Rep. {\bf 5}, 12295 (2015).

\bibitem{Zhou2015-2}
X. Zhou, W.-K. Lou, F. Zhai, and K. Chang, 
Anomalous magneto-optical response of black 
phosphorus thin films, 
Phys. Rev. B {\bf 92}, 165405 (2015).

\bibitem{Peeters}
M. Tahir, P. Vasilopoulos, and F. M. Peeters, 
Magneto-optical transport properties of 
monolayer phosphorene, 
Phys. Rev. B {\bf 92}, 045420 (2015).

\bibitem{Low}
Y. Jiang, R. Roldan, F. Guinea, and T. Low, 
Magnetoelectronic properties of multilayer 
black phosphorus, 
Phys. Rev. B {\bf 92}, 085408 (2015).

\bibitem{Pereira}
J. M. Pereira, Jr. and M. I. Katsnelson, 
Landau levels of single-layer and bilayer 
phosphorene, 
Phys. Rev. B {\bf 92}, 075437 (2015). 

\bibitem{Zhou2017}
X. Zhou, W.-K. Lou, D. Zhang, F. Cheng, 
G. Zhou, and K. Chang, 
Effective $g$ factor in black phosphorus thin films, 
Phys. Rev. B {\bf 95}, 045408 (2017). 

\bibitem{Zhou2023}
P. Wu, Z.-G. Shi, X. Chen, and X. Zhou, 
Anisotropic magneto-optical transport properties 
in black phosphorus induced by in-plane magnetic field, 
J. Phys.: Condens. Matter {\bf 35}, 065701 (2023).

\bibitem{pump-probe}
X. Liu, W. Lu, X. Zhou, Y. Zhou, C. Zhang, J. Lai, 
S. Ge, M.C. Sekhar, S. Jia, K. Chang, and D. Sun, 
Dynamical anisotropic response of black phosphorus 
under magnetic field, 
2D Mater. {\bf 5}, 025010 (2018).

\bibitem{perucchi1}
P. Di. Pietro, M. Mitrano, S. Caramazza, F. Capitani, 
S. Lupi, P. Postorino, F. Ripanti, B. Joseph, N. Ehlen, 
A. Gruneis, A. Sanna, G. Profeta, P. Dore, and A. Perucchi, 
Emergent Dirac carriers across a pressure-induced 
Lifshitz transition in black phosphorus 
Phys. Rev. B {\bf 98}, 165111 (2018). 

\bibitem{layer-pressure}
S. Huang, Y. Lu, F. Wang, Y. Lei, C. Song, J. Zhang, 
Q. Xing, C. Wang, Y. Xie, L. Mu, G. Zhang, H. Yan, 
B. Chen, and H. Yan.
Layer-Dependent Pressure Effect on the Electronic 
Structure of 2D Black Phosphorus, 
Phys. Rev. Lett. {\bf 127}, 186401 (2021). 

\bibitem{goodenough}
For the development of high-pressure research 
on BP, see, for example, X. Li, J. Sun, P. Shahi, 
M. Gao, A. H. MacDonald, Y. Uwatoko, T. Xiang, 
J. B. Goodenough, J. Cheng, and J. Zhou, 
Pressure-induced phase transitions and superconductivity 
in a black phosphorus single crystal, 
Proc. Nat. Acad. Sci.{\bf 115}, 9935 (2018), 
and references cited therein.  

\bibitem{akiba}
K. Akiba, A. Miyake, Y. Akahama, K. Matsubayashi, 
Y. Uwatoko, and M. Tokunaga, 
Anomalous Quantum Transport Properties in 
Semimetallic Black Phosphorus, 
J. Phys. Soc. Jpn. {\bf 84}, 073708 (2015). 

\bibitem{xiang}
Z. J. Xiang, G. J. Ye, C. Shang, B. Lei, N. Z. Wang, 
K. S. Yang, D. Y. Liu, F. B. Meng, X. G. Luo, L. J. Zou, 
Z. Sun, Y. Zhang, and X. H. Chen, 
Pressure-Induced Electronic Transition in 
Black Phosphorus, 
Phys. Rev. Lett. {\bf 115}, 186403 (2015).  

\bibitem{mito} 
T. Fujii, Y. Nakai, M. Hirata, Y. Hasegawa, Y. Akahama, 
K. Ueda, T. Mito, 
Giant Density of States Enhancement Driven by a Zero-Mode 
Landau Level in Semimetallic Black Phosphorus under Pressure, 
Phys. Rev. Lett. {\bf 130}, 076401 (2023).

\bibitem{QHE}
See, for example, K. von Klitzing, 
The quantized Hall effect, 
Rev. Mod. Phys. {\bf 58}, 519 (1986). 

\bibitem{graphene-QHE-1}
K. S. Novoselov, A. K. Geim, S. V. Morozov, D. Jiang, 
M. I. Katsnelson, I. V. Grigorieva, S. V. Dubonos, 
and A. A. Firsov, 
Two-dimensional gas of massless Dirac fermions in 
graphene, 
Nature {\bf 438}, 197 (2005).

\bibitem{graphene-QHE-2}
Y. Zhang, Y.-W. Tan, H. L. Stormer, and P. Kim, 
Experimental observation of the quantum Hall effect 
and Berry's phase in graphene, 
Nature {\bf 438}, 201 (2005). 

\bibitem{CR}
See, for example, D. J. Hilton, T. Arikawa, 
and J. Kono, Cyclotron Resonance, in 
{\it Characterization of Materials, 2nd Ed.}, 
edited by E. N. Kaufmann (Wiley, 2012). 

\bibitem{heiman}
See, for example, D. Heiman, 
Laser Spectroscopy of Semiconductors at 
Low Temperatures and High Magnetic Fields, 
{\it Semiconductors and Semimetals Vol. 36}, 
edited by D. G. Seiler and C. L. Littler 
(Academic Press, Boston, MA, 1992), Chap. 1. 


\bibitem{Bi-1}
M. P. Vecchi, J. R. Pereira, and M. S. Dresselhaus, 
Anomalies in the magnetoreflection spectrum of 
bismuth in the low-quantum-number limit, 
Phys. Rev. B {\bf 14}, 298 (1976).

\bibitem{Bi-2}
A. A. Schafgans, K. W. Post, A. A. Taskin, Y. Ando, 
X.-L. Qi, B. C. Chapler, and D. N. Basov, 
Landau level spectroscopy of surface states in the 
topological insulator Bi$_{0.91}$Sb$_{0.09}$ via 
magneto-optics, 
Phys. Rev. B {\bf 85}, 195440 (2012). 

\bibitem{graphite1}
W. W. Toy, C. R. Hewes, and M. S. Dresselhaus, 
Magnetoreflection studies of single crystal 
pyrolytic and kish graphite, 
Carbon {\bf 11}, 575 (1973).  

\bibitem{graphite2}
Z. Q. Li, S.-W. Tsai, W. J. Padilla, S. V. Dordevic, 
K. S. Burch, Y. J. Wang, and D. N. Basov, 
Infrared probe of the anomalous magnetotransport 
of highly oriented pyrolytic graphite in the 
extreme quantum limit, 
Phys. Rev. B {\bf 74}, 195404 (2006). 

\bibitem{graphene-LL}
M. L. Sadowski, G. Martinez, M. Potemski, C. Berger, 
and W. A. de Heer, 
Landau Level Spectroscopy of Ultrathin Graphite Layers, 
Phys. Rev. Lett. {\bf 97}, 266405 (2006).

\bibitem{carrier}
The samples were not intentionally doped, but 
were naturally $p$-type with a carrier density 
of $(2$$-$$5) \times 10^{15}$~cm$^{-3}$ as evaluated 
by Hall effect measurement.  This density is too 
low to affect our $R(\omega)$ data, and the sample 
can be regarded as undoped.  

\bibitem{MO}
S. Kimura, T. Nishi, T. Takahashi, T. Hirono, 
Y. Ikemoto, T. Moriwaki, and H. Kimura, 
Infrared spectroscopy under extreme conditions, 
Physica B {\bf 329-333}, 1625 (2003).  

\bibitem{bl43ir}
T. Moriwaki and Y. Ikemoto, 
BL43IR at SPring-8 redirected, 
Infrared Phys. Tech. {\bf 51}, 400 (2008). 

\bibitem{IRSR-review}
S. Kimura and H. Okamura, 
Infrared and Terahertz Spectroscopy of Strongly 
Correlated Electron Systems under Extreme Conditions, 
J. Phys. Soc. Jpn. {\bf 82}, 021004 (2013).  


\bibitem{SI}
See the Supplemental Material at (URL to be inserted).   

\bibitem{mybook}
H. Okamura, Methods for obtaining the optical 
constants of a material, 
{\it Optical Techniques for Solid State Materials 
Characterization} edited by R. Prasankumar and 
A. Taylor (CRC, Boca Raton, FL, 2011), Chap. 4. 

\bibitem{footnote2}
To perform KK analysis on a measured $R(\omega)$, 
both ends of $R(\omega)$ were extrapolated as 
follows.  
A constant was smoothly connected to $R(\omega)$ 
below 265~meV, since strong periodic oscillations 
were present there due to multiple internal 
reflections in the sample.  Above 380~meV, 
an extrapolation was made following the actually 
measured $R(\omega)$ in visible and uv ranges 
\cite{morita}.  
Note that specific details of these 
extrapolations have only minor effects on 
the obtained $\sigma(\omega)$ and do not alter 
the main results of the present work.  

\bibitem{miura}
See, for example, N. Miura, Physics of 
Semiconductors in High Magnetic Fields 
(Oxford University Press, Oxford, 2008). 


\bibitem{gap-Tdep}
C. E. P. Villegas A. R. Rocha, and A. Marini, Anomalous 
Temperature Dependence of the Band Gap in Black 
Phosphorus, Nano Lett. {\bf 16}, 5095 (2016).

\bibitem{splitting}
Some of the spectra at highest fields in Fig.~2 exhibit two 
subpeaks for one $n$ state ({\it e.g.}, $n=3$ and 5 at 12~T) 
and Y peak also seems to show fine splitting at 12~T.  
They may have actually resulted from Zeeman splitting of 
$p$-derived states at high $B$, although their 
exact origin is unclear.

\bibitem{footnote4}
These two transitions, corresponding to 
$\Delta m_j = \pm 1$, are allowed with left and 
right circular polarizations.  In our experiment, 
the incident light had a linear polarization, 
which can be regarded as a superposition of 
the two circular polarizations.  
Therefore, both these transitions may be 
observed in our experiment.  

\bibitem{perovskite}
See, for a recent example, Y. Yamada, H. Mino, 
T. Kawahara, K. Oto, H. Suzuura, and Y. Kanemitsu, 
Polaron masses in CH$_3$NH$_3$Pb$X_3$ perovskites 
determined by Landau level spectroscopy in low 
magnetic fields, 
Phys. Rev. Lett. {\bf 126}, 237401 (2021).  

\bibitem{BP-exciton}
E. Carre, L. Sponza, A. Lusson, I. Stenger, E. Gaufres, 
A Loiseau, and J. Barjon, 
Excitons in bulk black phosphorus evidenced by 
photoluminescence at low temperature, 
2D Materials {\bf 8}, 021001 (2021).  

\bibitem{Zhou2024}
X. Zhou, unpublished work and private communications.

\bibitem{footnote3}
The results of the quadratic fitting are the following.  
For X peak, $a=4.74 \times 10^{-3}$~meV/T$^2$, 
$b=3.21 \times 10^{-2}$~meV/T with $E_\text{X}$=273.0~meV.  
For Y peak, $a=9.26 \times 10^{-3}$~meV/T$^2$, 
$b=1.08 \times 10^{-1}$~meV/T with $E_\text{Y}$=279.3~meV.  

\bibitem{radius}
For a hydrogenic (isotropic) exciton, 
$E_b = \hbar^2 / (2 m_r^\ast {a^\ast}^2)$ \cite{miura}.  
Substituting $E_b=9.7$~meV and $m_r^\ast=0.129 m_0$ 
into this equation yields $a^\ast=55~\AA$.  
Note that a fully anisotropic theoretical model 
in Ref.~\cite{BP-exciton} predicts exciton 
extensions of 90.9~$\AA$, 39.6~$\AA$, and 63.1~$\AA$ 
along the $x$, $y$, and $z$ axes, respectively.  
Their average is 65~$\AA$ and about 
20~$\%$ larger than our rough estimate above.  

\bibitem{diamagnetic}
R. P. Seisyan, 
Diamagnetic excitons and exciton magnetopolaritons 
in semiconductors, 
Semicond. Sci. Technol. {\bf 27}, 053001 (2012).  

\bibitem{exciton}
For a review on 1$s$ and 2$s$ exciton absorption in 
2D and 3D semiconductors, see, for example, sections 
2 and 3.1 in K. R. Hansen, J. S. Colton, and 
L. Whittaker-Brooks, 
Measuring the exciton binding energy: Learning 
from a decade of measurements on halide perovskites 
and transition metal dichalcogenides, 
Adv. Opt. Mater. {\bf 12}, 2301659 (2024).

\bibitem{okamura2}
H. Okamura, D. Heiman, M. Sundaram, and A. C. Gossard, 
Inhibited recombination of charged magnetoexcitons, 
Phys. Rev. B {\bf 58}, R15985 (1998).


\end{thebibliography}
\end{document}